\documentclass[osajnl,preprint,showpacs,superscriptaddress,12pt]{revtex4-1} 

\usepackage{amsmath,amssymb,graphicx}

\begin{document}

\title{Ermakov-Lewis symmetry in photonic lattices }

\author{B. M. Rodr\'{\i}guez-Lara }
\email{bmlara@inaoep.mx}
\affiliation{Instituto Nacional de Astrof\'{i}sica, \'{O}ptica y Electr\'{o}nica \\ Calle Luis Enrique Erro No. 1, Sta. Ma. Tonantzintla, Pue. CP 72840, M\'{e}xico}

\author{P. Aleahmad}
\affiliation{CREOL/College of Optics and Photonics, University of Central Florida, Orlando, Florida, USA.}

\author{H. M. Moya-Cessa}
\affiliation{Instituto Nacional de Astrof\'{i}sica, \'{O}ptica y Electr\'{o}nica \\ Calle Luis Enrique Erro No. 1, Sta. Ma. Tonantzintla, Pue. CP 72840, M\'{e}xico}

\author{D. N. Christodoulides}
\affiliation{CREOL/College of Optics and Photonics, University of Central Florida, Orlando, Florida, USA.}

\begin{abstract}
We present a class of waveguide arrays that is the classical analog of a quantum harmonic oscillator where the mass and frequency depend on the propagation distance.
In these photonic lattices refractive indices and second neighbor couplings define the mass and frequency of the analog quantum oscillator, while first neighbor couplings are a free parameter to adjust the model.
The quantum model conserves the Ermakov-Lewis invariant, thus the photonic crystal also posses this symmetry.

\end{abstract}


\maketitle 


Symmetry in classical and quantum mechanics provides a tool to gain insight from complex systems \cite{Boyer1974p589,Boyer1976p35,Leviathan2011p93}.
In its more pure form, abstract invariants are related to physical conservation laws as found by Noether \cite{Noether1918p235}.
In practice, the sets of transformations related to a symmetry help us making tractable an otherwise complex-in-appearance model \cite{Boyer1974p589}.
Here we are interested in such an optical system, a specific class of photonic lattices \cite{Christodoulides2003p817} composed of waveguides that interact with their first and second nearest neighbors through evanescent coupling.
Furthermore, the refractive indices and separation between waveguides are functions of the propagation distance.
In the following, we will show that such a class of photonic lattices is the classical analog of a propagation-dependent harmonic oscillator with an Ermakov-Lewis invariant (ELI) \cite{Ermakov1880p1,Lewis1967p510,Lewis1968p1976}.
This relation to the time-dependent harmonic oscillator makes these photonic crystals a suitable candidate for the classical simulation of diverse quantum problems; e.g., a particle moving in a magnetic field or under the presence of friction.
Finally, we will provide an analytic propagator for classical light impinging these arrays of coupled waveguides through the symmetry transformations related to the ELI.


Our photonic lattice is semi-infinite and composed by individual waveguides whose refractive index varies linearly with their position in the array times a common function of the propagation distance.
The couplings between first and second neighbor waveguides vary as the square root of a function of the position in the array times a common function of the propagation distance; there is one particular function for first neighbors and another for second neighbors.
In short, it is described by the differential set
\begin{eqnarray}
 i \partial_{z} \mathcal{E}_{n}(z) + \alpha_{0}(z) ~n ~\mathcal{E}_{n}(z) +  \alpha_{1}(z) && \left[ f_{n+1} ~\mathcal{E}_{n+1}(z) + f_{n} ~\mathcal{E}_{n-1}(z) \right] +  \nonumber \\
&& + \alpha_{2}(z) \left[ g_{n+2} ~\mathcal{E}_{n+2}(z) + g_{n} ~\mathcal{E}_{n-2}(z) \right] =0
\end{eqnarray}
with $f_{n}= \sqrt{n}$, $g_{n} = \sqrt{n(n-1)}$ and $\mathcal{E}_{- \vert n \vert}(z)=0$, where the field amplitude at the $n$th waveguide is $\mathcal{E}_{n}(z)$, the common $z$-dependent refractive index is given by $\alpha_{0}(z)$ and the first and second neighbor coupling functions by $\alpha_{1}(z)$ and $\alpha_{2}(z)$, respectively.
The corresponding array of waveguides would not necessary be one-dimensional in order to fulfill the relations for first and second neighbor couplings.
If we define a state vector, $\vert \psi(z) \rangle = \sum_{j=0}^{\infty} \mathcal{E}_{j}(z) \vert j \rangle$, via the field amplitudes at the $j$th waveguides, $\mathcal{E}_{j}$, the differential set can be written as a Schr\"odinger-like equation, 
\begin{eqnarray}
i \partial_{z} \vert \psi(z) \rangle = \hat{H} \vert \psi(z) \rangle, \label{eq:SchEq}
\end{eqnarray}
in terms of creation (annihilation) operators, $\hat{a}^{\dagger}$ ($\hat{a}$),
\begin{eqnarray}
\hat{H} &=&  - \left[ \alpha_{0}(z) \hat{n} + \alpha_{1}(z) \left( \hat{a} + \hat{a}^{\dagger}\right) + \alpha_{2}(z) \left(\hat{a}^{2} + \hat{a}^{\dagger 2}\right)\right].
\end{eqnarray}
The action of the bosonic creation (annihilation) and number operators are given by $\hat{a}^{\dagger} \vert n \rangle = \sqrt{n+1} \vert n+1 \rangle$ ($\hat{a} \vert n \rangle = \sqrt{n} \vert n \rangle$) and $\hat{n} \vert n \rangle = n \vert n \rangle$, in that order.
At this point we can move to normalized canonical position and momentum, $\hat{a} = \frac{1}{\sqrt{2}} \left( \hat{q} + i \hat{p} \right)$ and $\hat{a}^{\dagger} = \frac{1}{\sqrt{2}} \left( \hat{q} - i \hat{p} \right)$, to rewrite the Hamiltonian as
\begin{eqnarray}
\hat{H} &=& - \left[ \frac{1}{2 M(z)} \hat{p}^{2} + \frac{1}{2} M(z) \Omega^{2}(z) \hat{q}^{2} +  \sqrt{2} \alpha_{1}(z) \hat{q} - \frac{\alpha_{0}(z)}{2} \right], \label{eq:Osc}
\end{eqnarray}
where the $z$-dependent mass and frequency are 
\begin{eqnarray}
M(z) &=& \frac{1}{ \alpha_{0}(z) - 2 \alpha_{2}(z)}, \\
\Omega^{2}(z) &=& \alpha_{0}^{2}(z) - 4 \alpha_{2}^{2}(z).
\end{eqnarray}
Note that we have the restriction $\alpha_{0}(z)  \neq 2 \alpha_{2}(z) $ between the $z$-dependent refractive index and second neighbor couplings in order to work with a well behaved mass function.
In this form, a displacement and overall phase help us simplifying the dynamics, 
\begin{eqnarray}
\vert \psi(z) \rangle = e^{- i \int \varphi(z) dz } e^{ - i \left[ u(z) \hat{p} + M(z) \dot{u}(z) \hat{q}\right]} \vert \xi(z) \rangle, 
\end{eqnarray}
with $\varphi(z) = \left[\alpha_{0}(z) + M(z) \dot{u}^{2}(z)+ M(z) \Omega^{2}(z) u^2(z) \right]/2$ and the auxiliary function $u(z)$ such that it fulfills $\ddot{u}(z) + \dot{M}(z) \dot{u}(z) / M(z) + \Omega^{2}(z) u + \sqrt{2} \alpha_{1}(z)/M(z) =0$, where we have used the shorthand notation  $\dot{u}(z) = \partial_{z} u(z)$.
This is a good point to stop and note that the first coupling neighbor $\alpha_{1}(z)$ influences the propagation in our photonic lattices only through its role in defining the auxiliary function $u(z)$.
Thus, we can rewrite Eq. (\ref{eq:SchEq}) as a harmonic oscillator where the mass and frequency depend on the propagation distance, 
\begin{eqnarray}
i \partial_{t} \vert \xi(t) \rangle =  \left[ \frac{1}{2 m(t)} \hat{p}^{2} + \frac{1}{2} m(t) \omega^{2}(t) \hat{q}^{2} \right]  \vert \xi(t) \rangle,
\end{eqnarray}
and we have made the variable change $t=-z$ leading to $m(t)=M(-t)$ and $\omega(t) = \Omega(-t)$ for the sake of simplicity.
Such a quantum model shows a Lewis-Ermakov invariant \cite{Lewis1967p510,Lewis1968p1976,Lewis1969p1458,Eliezer1976p463,Lutzky1978p3,Pedrosa1997p3219},
\begin{eqnarray}
\hat{I} = \frac{1}{2} \left\{ \left[ \frac{\hat{q}}{\rho(t)}\right]^{2} + \left[ \rho(t) \hat{p} - m(t) \dot{\rho}(t) \hat{q} \right]^{2} \right\},
\end{eqnarray}
where the new auxiliary function fulfills Ermakov equation \cite{Ermakov1880p1},
\begin{eqnarray}
\ddot{\rho}(t) + \frac{\dot{m}(t)}{m(t)} \dot{\rho}(t) + \omega^{2}(t) \rho(t) = \frac{1}{m^{2}(t) \rho^{3}(t)}.
\end{eqnarray}
The Lewis-Ermakov invariant posses a related set of symmetry transformations  \cite{Pedrosa1997p3219,FernandezGuasti2003p2069},
\begin{eqnarray}
\vert \xi(t) \rangle &=& e^{ i \frac{m(t) \dot{\rho}(t) }{2 \rho(t)} \hat{q}^{2}}  e^{-i \frac{\ln \rho(t)}{2} \left( \hat{p} \hat{q} + \hat{q} \hat{p} \right) } \vert \zeta(t) \rangle,
\end{eqnarray}
which are equivalent to displacement, first exponential in rhs term, and squeezing operations, second exponential in rhs term.
This squeezed and displaced basis diagonalizes our system:
\begin{eqnarray}
i \partial_{t} \vert \zeta (t)\rangle &=& \frac{1}{2 m(t) \rho^{2}(t)} \left( \hat{p}^{2} + \hat{q}^{2} \right)\vert \zeta (t)\rangle, \\
&=&\frac{1}{ m(t) \rho^{2}(t)} \left( \hat{a}^{\dagger} \hat{a} + \frac{1}{2} \right)\vert \zeta(t) \rangle.
\end{eqnarray}
It is straightforward to solve this Sch\"odinger-like equation as all the dependence on the propagation has been factorized to a common term.
Thus, the analytic propagator for any given initial state is
\begin{eqnarray}
\vert \zeta(t) \rangle &=& e^{-i \int \frac{1}{m(t) \rho^{2}(t)} dt \left( \hat{a}^{\dagger} \hat{a} + \frac{1}{2} \right) } \vert \zeta(0) \rangle,
\end{eqnarray}
where the state vector $\vert \zeta(0) \rangle$ holds the information from the initial field amplitudes that impinge into the waveguide array.
These transformations and evolution in the new frame of reference are enough to provide an impulse function in the original frame.
Notice that light impinging just the $j$th waveguide in the original frame of reference will be equivalent to having the initial wavefunction $\frac{1}{\sqrt{2^{j} j!}} \left( \frac{1}{\pi} \right)^{1/4} e^{-q^2/2} H_{j}(q)$ where $H_{n}(x)$ is the $n$th Hermite polynomial \cite{Abramowitz1970}.
Thus, a beam of light impinging just the first waveguide of the array is equivalent to an initial Gaussian wavefunction in dimensionless canonical space, $\Psi_{0}(q) = e^{-q^{2}/2} / \pi^{1/4}$. 
This has to be taken into account when going to/from the transformed frame.

Furthermore, as our photonic lattice with first and second neighbor coupling is equivalent to a $z$-dependent harmonic-oscillator, Eq. (\ref{eq:Osc}), it can be used to simulate a charged particle moving nonrelativistically in the presence of a magnetic field \cite{Lewis1967p510,Lewis1969p1458} if the parameters are adjusted such that the mass is constant, $m(t) = m$, and the frequency is proportional to the amplitude of the magnetic field, $\omega(t) = B(t) / 2$.
It can also be used to simulate quantum oscillators in the presence of friction if the mass is written as $m(t)= m / F(t)$ \cite{Pedrosa1987p2662}; setting $F(t)=e^{- \gamma t}$ with constant $\gamma$ and $\omega(t)$ leads to the Caldirola-Kanai Hamiltonian \cite{Caldirola1941p393,Kanai1948p440}.
Analytical closed forms for the auxiliary function $\rho(t)$ can be obtained for a variety of driving frequencies; e.g. $\omega(t) = c$ leads to $\rho^{2}(t) = \omega(t)^{-1}$, $\omega(t) = c t^{-1}$ gives $\rho^{2}(t) = t \left( c^{2} - 1/4 \right)^{-1/2}$, $\omega(t)= c t^{k}$ yields $\rho^{2}(t) = \pi (k+1) t \left[ J_{(k+1)/2)^{2}} (k+1 c^{2} t^{k+1}) + Y_{(k+1)/2)^{2}} (k+1 c^{2} t^{k+1}) \right]/2$,  in all cases $c$ is a constant \cite{Lewis1968p1976,Eliezer1976p463}.
Of course, each case provides its own set of design challenges in order to bring the photonic analogues to the laboratory. 

\begin{figure}[htbp]
\centerline{\includegraphics[scale=1]{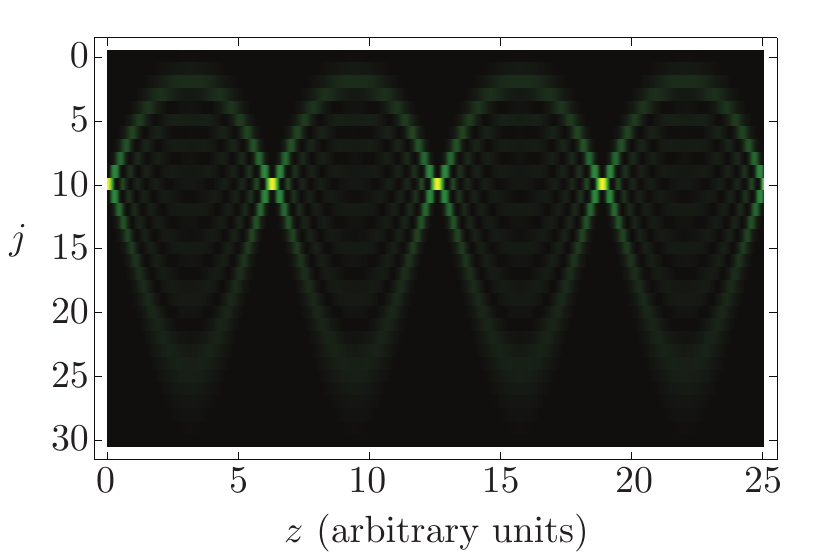}}
\caption{(Color Online) Light intensity propagation for a beam impinging the $j=10$ waveguide of a lattice with parameters $\alpha_{0}(z) = \alpha_{1}(z)=1$ and $\alpha_{2}(z)=0$ equivalent to an oscillator with constant mass and frequency, $M(z)=\Omega(z)=1$.}\label{fig:Fig1}
\end{figure}

\begin{figure}[htbp]
\centerline{\includegraphics[scale=1]{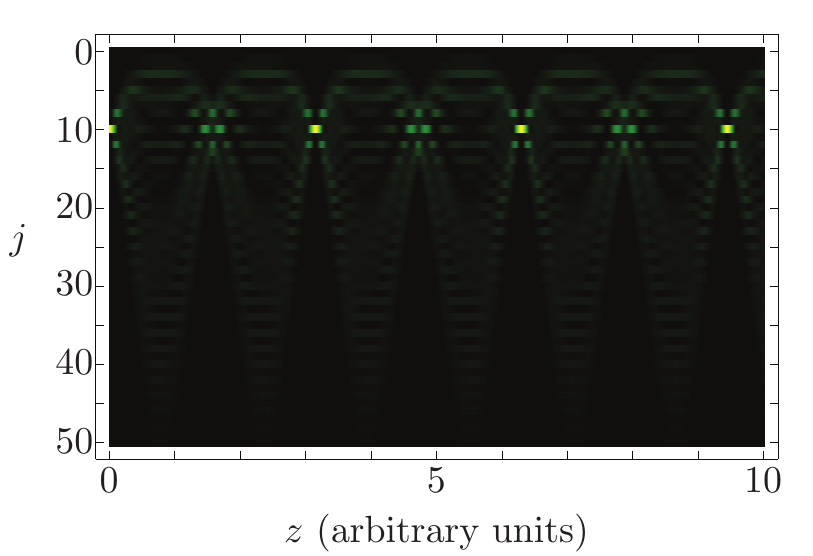}}
\caption{(Color Online) Light intensity propagation for a beam impinging the $j=10$ waveguide of a lattice with parameters $\alpha_{0}(z) = 5/2$, $\alpha_{1}(z)=1$ and $\alpha_{2}(z)=3/4$ equivalent to an oscillator with constant mass and frequency, $M(z)=1$ and $\Omega(z)=2$.}\label{fig:Fig2}
\end{figure}

Let us consider a simple example by setting an harmonic oscillator with constant mass and frequency, $M(z)=\Omega(z)=1$. 
This case is related to an array with parameters $\alpha_{0}(z) = 1$, $\alpha_{1}(z) = 1$ and $\alpha_{2}(z) = 0$; i.e, there are only first neighbor couplings in the photonic lattice.
This realization of our lattice is equivalent up to a constant with the Glauber-Fock oscillator lattice that allows for Block-like revivals and has been produced experimentally \cite{Keil2012p3801}. 
Figure \ref{fig:Fig1} shows the propagation of light intensity in such a lattice for an initial field impinging just the $j=10$ waveguide.
Increasing the constant driving frequency in the quantum model is equivalent to include second neighbor couplings in the optical model.
As the quantum models are the same, the dynamics of light propagating through the new waveguide array are similar to that of the Glauber-Fock oscillator lattice; e.g., the spatial frequency of the Bloch-like revivals becomes higher in the lattice including second neighbor couplings compared to that including just the first neighbor couplings; e.g. Figure \ref{fig:Fig2} shows propagation in our lattice with parameters $\alpha_{0}(z) = 5/2$, $\alpha_{1}(z)=1$ and $\alpha_{2}(z)=3/4$ which are related to the oscillator parameters $M(z)=1$ and $\Omega(z)=2$.

\begin{figure}[htbp]
\centerline{\includegraphics[scale=1]{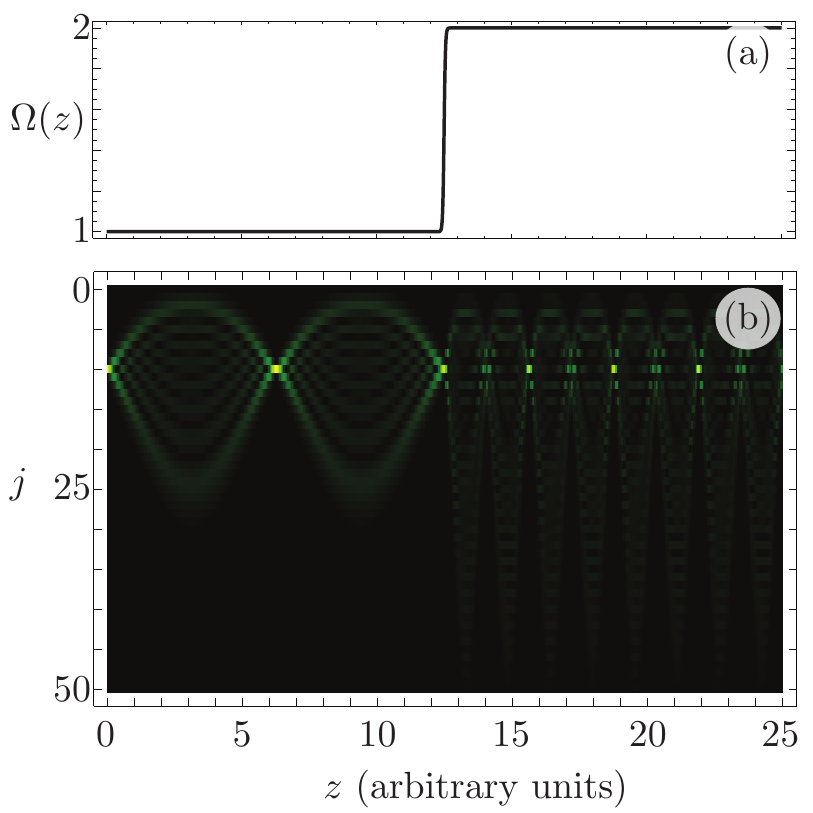}}
\caption{(Color Online)  (a) Profile of the driving frequency $\Omega(z)$. (b) Light intensity propagation for a beam impinging the $j=10$ waveguide of a lattice with parameters $\alpha_{0}(z)= \left[ M^2(z) \Omega(z)^{2} +1 \right]/\left[2 M(z)\right] $, $\alpha_{1}(z)=1$ and $\alpha_{2}(z)=\left[ M^2(z) \Omega(z)^{2} -1 \right]/\left[4 M(z) \right]$ with harmonic oscillator parameters $M(z)=1$ and $\Omega(z) = \left[3 + \tanh 20(z-12.5) \right]/2$.} \label{fig:Fig3}
\end{figure}

We can also study an example bridging the two arrays of waveguides discussed above by   considering a driving frequency $\Omega(z) = \left[3 + \tanh \epsilon (z-z_{s})\right]/2$ that is the equivalent of a smooth, well-behaved step function where the parameter $\epsilon$ controls the steepness of the jump and $z_{s}$ is the switching position, Fig. \ref{fig:Fig3}(a). 
Thus for propagation distances $z < z_{s}$ we expect the coherent oscillations related to the Glauber-Fock oscillator lattice with a given spatial frequency and then, after the switching distance $z_{s}$, we expect oscillations related to the higher spatial frequency of a Glauber-Fock oscillator including second neighbor couplings. 
Figure \ref{fig:Fig3}(b) shows the numerical propagation of such a system. 
This case admits an approximate auxiliary function $\rho^{2}(t) =  1 + 1/ \omega^{2}(t) + \left( 1 - 1/ \omega^{2}(t)  \right) \cos  \int_{t_{s}}^{t} 2\omega(\zeta) d\zeta  $ \cite{MoyaCessa2003p1,Guasti2004p188}.
As mentioned before this is equivalent to a Glauber-Fock oscillator lattice that smoothly transitions from just first neighbor couplings to first and second neighbor couplings.
The numerical simulations where carried with lattices consisting of 500 waveguides, the light intensity at the last waveguide was never larger than $10^{-6}$.

In short, we have shown that certain nontrivial arrays of photonic waveguides where refractive indices, first and second neighbor couplings depend on the propagation distance can classically simulate the dynamics of a quantum harmonic oscillator with non-constant mass and frequency.
The quantum system shows an Ermakov-Lewis invariant that defines the underlying symmetry of the photonic crystal and is related to diverse quantum mechanical problems; e.g. a charged particle in the presence of a magnetic field and driving/friction in quantum oscillators.

\bibliographystyle{osajnl}

\end{document}